\documentclass[runningheads]{llncs}
\usepackage{graphicx}
\usepackage{cite}

\usepackage{booktabs}
\usepackage{multirow}
\usepackage{makecell}
\usepackage{tabularx}
\usepackage{subfig}

\begin{document}
\title{MRI Brain Tumor Segmentation and Uncertainty Estimation using 3D-UNet architectures}

\author{Laura Mora Ballestar \and Veronica Vilaplana
\thanks{This work has been partially supported by the project MALEGRA TEC2016-75976-R financed by the Spanish Ministerio de Econom\'{i}a y Competitividad.} }

\authorrunning{L. Mora et al.}
%
\institute{Signal Theory and Communications Department, Universitat Politècnica de Catalunya. BarcelonaTech, Spain \\
\email{lmoraballestar@gmail.com, veronica.vilaplana@upc.edu}\\
}

\titlerunning{MRI Brain Tumor Segmentation and Uncertainty Estimation using 3D-UNet}
\maketitle  
\begin{abstract}
\textit{Automation of brain tumor segmentation in 3D magnetic resonance images (MRIs) is key to assess the diagnostic and treatment of the disease. In recent years, convolutional neural networks (CNNs) have shown improved results in the task. However, high memory consumption is still a problem in 3D-CNNs. Moreover, most methods do not include uncertainty information, which is especially critical in medical diagnosis. This work studies 3D encoder-decoder architectures trained with patch-based techniques to reduce memory consumption and decrease the effect of unbalanced data. The different trained models are then used to create an ensemble that leverages the properties of each model, thus increasing the performance. We also introduce voxel-wise uncertainty information, both epistemic and aleatoric using test-time dropout (TTD) and data-augmentation (TTA) respectively. In addition, a hybrid approach is proposed that helps increase the accuracy of the segmentation. The model and uncertainty estimation measurements proposed in this work have been used in the BraTS'20 Challenge for task 1 and 3 regarding tumor segmentation and uncertainty estimation.}

\end{abstract}

\section{Introduction}

Brain tumors are categorized into primary, brain originated; and secondary, tumors that have spread from elsewhere and are known as brain metastasis tumors. Among malignant primary tumors, gliomas are the most common in adults, representing 81\% of brain tumors \cite{epidemiology}. The World Health Organization (WHO) categorizes gliomas into grades I-IV which can be simplified into two types (1) “low grade gliomas” (LGG), grades I-II, which are less common and are characterized by low blood concentration and slow growth and (2) “high grade gliomas” (HGG), grades III-IV, which have a faster growth rate and aggressiveness.

The extend of the disease is composed of four heterogeneous histological sub-regions, i.e. the peritumoral edematous/invaded tissue, the necrotic core (fluid-filled), the enhancing and non-enhancing tumor (solid) core. Each region is described by varying intensity profiles across MRI modalities (T1-weighted, post-contrast T1-weighted, T2-weighted, and Fluid-Attenuated Inversion Recovery-FLAIR), which reflect the diverse tumor biological properties and are commonly used to assess the diagnosis, treatment and evaluation of the disease. These MRI modalities facilitate tumor analysis, but at the expense of performing manual delineation of the tumor regions which is a challenging and time-consuming process. For this reason, automatic mechanisms for region tumor segmentation have appeared in the last decade thanks to the advancement of deep learning models in computer vision tasks. Despite these recent advances, the segmentation of brain tumors in multimodal MRI scans is still a challenging task in medical image analysis due to the highly heterogeneous appearance and shape of the problem.

The Brain Tumor Segmentation (BraTS) \cite{brats, advancing_brats, bakas_identifying, segmentation_label_gbm, segmentation_label_lgg} challenge started in 2012 with a focus on evaluating state-of-the-art methods for glioma segmentation in multi-modal MRI scans. BraTS 2020 training dataset includes 369 cases (293 HGG and 76 LGG), each with four 3D MRI modalities rigidly aligned, re-sampled to $1 mm^3$ isotropic resolution and skull-stripped with size $240x240x155$. Each provides a manual segmentation approved by experienced neuro-radiologists. Training annotations comprise the enhancing tumor (ET, label 4), the peritumoral edema (ED, label 2), and the necrotic and non-enhancing tumor core (NCR/NET, label 1). The nested sub-regions considered for evaluation are: whole tumor WT (label 1, 2, 4), tumor core TC (label 1, 4) and enhancing tumor ET (label 4). The validation set includes 125 cases, with unknown grade nor ground truth annotation. The test set is composed of 166 cases.

The goal of this work is to develop a 3D convolutional neural network (CNN) for brain tumor segmentation from 3D MRIs and provide an uncertainty measure to assess the confidence on the model predictions. The proposed methods are used to participate in BraTS'20 Challenge for tasks 1 and 3, respectively. In task 1, we explore the use of two well-known 3D-CNN for medical imaging --V-Net \cite{vnet} and 3D-UNet \cite{3dunet}-- and apply some modifications to their baselines. With both networks, the usage of sampling techniques is necessary due memory limitations as well as data augmentation to prevent over-fitting. For task 3, the work provides voxel-wise uncertainty measures computed at test time, with global and per sub-region information. Uncertainty is estimated using both epistemic and aleatoric \cite{unc_categorization} uncertainties using test-time dropout (TTD) \cite{montecarlo_dropout} and data augmentation, respectively.
\section{Related Work}

\subsection{Semantic Segmentation}
Brain tumor segmentation methods include generative and discriminative approaches. Generative methods try to incorporate prior knowledge and model probabilistic distributions whereas discriminative methods extract features from image representations. This latter approach has thrived in recent years thanks to the advancement in CNNs, as demonstrated in the winners of the previous BraTS. The biggest break through in this area was introduced by DeepMedic \cite{deepmedic} a 3D CNN that exploits multi-scale features using parallel pathways and incorporates a fully connected conditional random field (CRF) to remove false positives. \cite{Casamitjana1} compares the performances of three 3D CNN architectures showing the importance of the multi-resolution connections to obtain fine details in the segmentation of tumor sub-regions.
More recently, EMMA \cite{ensembles} creates and ensemble at inference time which reduces overfitting but at high computational cost, and \cite{Casamitjana2} proposes a cascade of two CNNs, where the first network produces raw tumor masks and the second network is trained on the vecinity of the tumor to predict tumor regions. BraTS 2018 winner \cite{myronenko20183d} proposed an asymmetrically encoder-decoder architecture with a variational autoencoder to reconstruct the image during training, which is used as a regularizer. Isensee, F \cite{no_newnet} uses a regular 3D-U-Net optimized on the evaluation metrics and co-trained with external data. BraTS 2019 winners \cite{two-stage} use a two-stage cascade U-Net trained end-to-end. Finally, \cite{tricks} applies several tricks in three categories: data processing, model devising and optimization process to boost the model performance.

\subsection{Uncertainty}

Uncertainty information of segmentation results is important, specially in medical imaging, to guide the clinical decisions and help understand the reliability of the provided segmentation, hence being able to identify more challenging cases which may require expert review. Segmentation models for brain tumor MRIs tend to label voxels with less confidence in the surrounding tissues of the segmentation targets \cite{wangUnc}, thus indicating regions that may have been miss-segmented.

Last year's BraTS challenge already started introducing uncertainty measurements. \cite{demistifying} computes epistemic uncertainty using TTD. They obtain a posterior distribution generated after running several epochs for each image at test-time. Then, mean and variance are used to evaluate the model uncertainty. A different approach is proposed by Wang G \cite{wangUnc}, who uses TTD and data augmentation to estimate the voxel-wise uncertainty by computing the entropy instead of the variance. Finally, \cite{McKinleyUnc} proposes to incorporate uncertainty measures during training as they define a loss function that models label noise and uncertainty.
\section{Method}

\subsection{Dataset Statistics}\label{data}

\begin{figure}[ht] 
    \centering
    \includegraphics[width=1\textwidth]{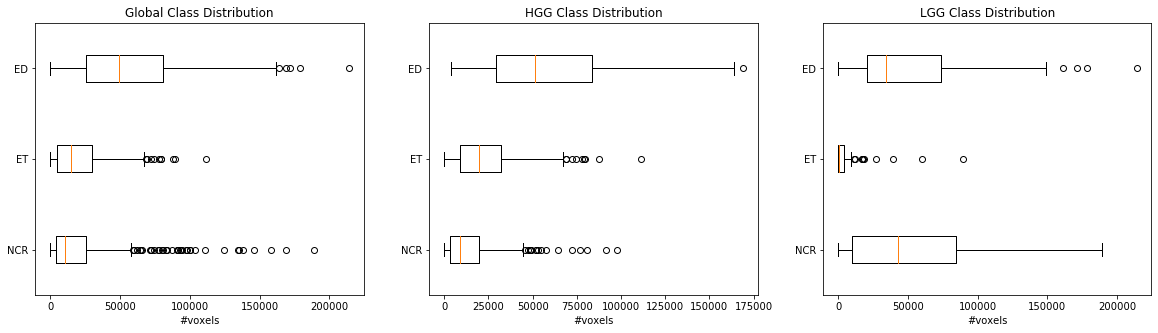}
    \caption{Distribution of each class ED, ET, NCR. From left to right, (1) number of voxels in all cases, (2) number of voxels for the HGG and (3) number of voxels for the LGG} 
    \label{fig:class-distribtuion}
\end{figure}

The biggest complexity for brain tumor segmentation is derived from the class imbalance. The tumor regions account for a 5-15\% of the brain tissue and each tumor region is an even smaller portion. Fig. \ref{fig:class-distribtuion} provides a graphical representation of the distribution per each tumor class: ET, NCR, ED; without healthy tissue. It can be seen, that ED is more probable than ET and NCR and that there is high variability between subjects in the NCR label. Another complexity is the difference between glioma grades as LGG patients are characterized by low blood concentration which is translated to low appearance of ET voxels and higher number of voxels for NCR and NET regions.

\subsection{Data Pre-processing and Augmentation}\label{pre-processing}

MRI intensity values are not standardized as the data is obtained from different institutions, scanners and protocols. Therefore we normalize each modality of each patient independently to have zero mean and unit std based on non-zero voxels only, which represent the brain region.

We also apply data augmentation techniques to prevent over-fitting by trying to disrupt minimally the data. For this, we apply Random Flip (for all 3 axes) with a 50\% probability, Random $90^{\circ}$ Rotation on two axis with a 50\% probability, Random Intensity Shift between ($-0.1..0.1$ of data std) and Random Intensity Scale on all input channels at range (0.9..1.1).

\subsection{Sampling Strategy} \label{sampling}
3D-CNNs are computationally expensive and in many cases, the input data cannot be fed directly to the network. Patch-wise training helps to free memory resources so more images can be fed in one batch. However, there is a trade-off between patch size and batch size. Bigger batches will have a more accurate representation of the data but will require smaller patches (due to memory constraints) that provide local information but lack contextual knowledge.

Another key aspect to consider when selecting the patching strategy is to maintain the class distribution. Losing this distribution can generate a biased model, i.e. if the model only sees small patches with tumor it will likely miss-classify healthy tissue.

In this work, we have used two approaches depending on the patch size. 

\begin{itemize}
    \item Binary Distribution: Small patches, equal or lower than $64^3$ are randomly selected with a 50\% probability of being centred on healthy tissue and 50\% probability on tumor \cite{deepmedic}.
    \item Random Tumor Distribution: Bigger sizes, $112^3$ or $128^3$, are selected randomly but always centred in tumor region, as the patches will contain more healthy tissue and background information.
\end{itemize}

\subsection{Loss} \label{loss}

The Dice score coefficient (DSC) is a measure of overlap widely used to assess segmentation performance when ground truth is available. Proposed in Milletari et al. \cite{vnet} as a loss function for binary classification, it can be written as:

\begin{equation}
L_{dice} =  1 - \frac{2*\sum_{i=1}^{N}p_{i}g_{i}}{\sum_{i}^{N}p_{i}^{2}+\sum_{=1i}^{N}g_{i}^{2} + \epsilon }
\end{equation}

where N is the number of voxels, $p_{i}$  and $g_{i}$ correspond to the predicted and ground truth labels per voxel respectively, and $\epsilon$ is added to avoid zero division.

Many variations of the dice loss have been proposed in the literature. For instance, the Generalized Dice Loss (GDL) \cite{gen_dice} which is based on the generalized dice score (GDS) \cite{crum_dice_score} for multiple class evaluation. Its goal is to correct the correlation between region size and dice score, by weighting the contribution of each label with the inverse of its volume. It is described as:

\begin{equation}
L_{diceGDL} =  1 - 2 \frac{\sum_{l=1}^{L} w_{l} \sum_{i=1}^{N} p_{li} g_{li}}{\sum_{l=1}^{L} w_{l} \sum_{i=1}^{N} p_{li} + g_{li} + \epsilon}
\end{equation}

where L represents the number of classes and $w\_{l}$ the weight given to each class. We use the GDL variant as it is more suited for unbalanced segmentation problems.

\subsection{Network Architecture}\label{network-arch}

This work proposes three networks, variations of V-Net \cite{vnet} and 3U-Net \cite{3dunet} architectures, for brain tumor segmentation and creates an ensemble to mitigate the bias in each independent model.

The different models are trained using the ADAM optimizer, with start learning rate of $1e-4$, decreased by a factor of 5 whenever the validation loss has not improved in the past 30 epochs and regularized with a l2 weight decay of $1e-5$. They all use the GDL loss.

\subsubsection{\textbf{V-Net}}

\begin{figure}[ht]
    \centering
    \includegraphics[width=1\textwidth]{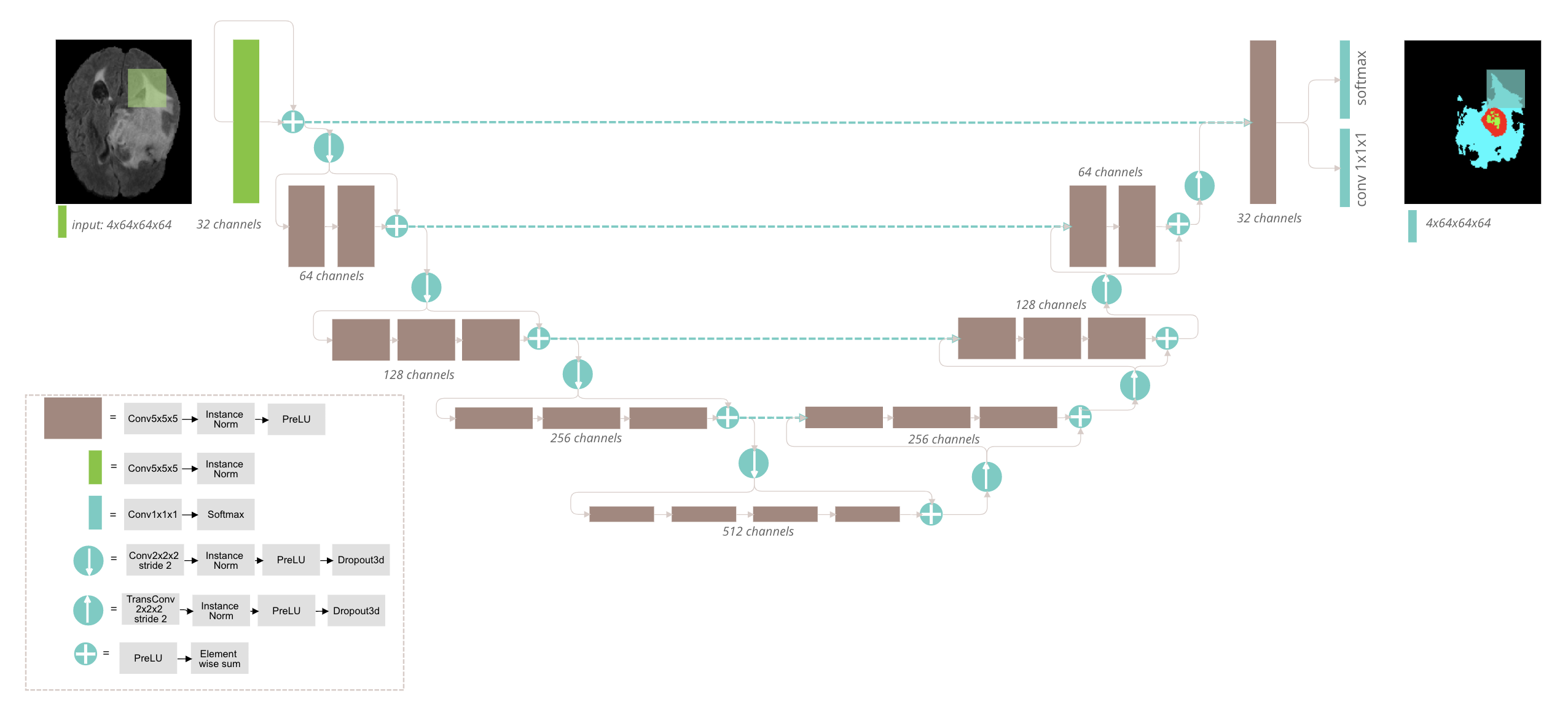}
    \caption{V-Net \cite{vnet} architecture with instance normalization, PreLU non-linearities, 32 feature channels at the highest resolution. Feature dimensionality is denoted at each block. The network outputs the segmentation and the softmax prediction.} 
    \label{fig:vnet}
\end{figure}

The V-Net implementation has been adapted to use four output channels (Non-Tumor, ED, NCR/NET, ET) and uses Instance Normalization \cite{ulyanov2016instance} in contrast to Batch Normalization, which normalizes across each channel for each training example instead of the whole batch. Also, as proposed in \cite{no_newnet}, we have increased the number of feature maps to 32 at the highest resolution, instead of 16 as proposed by the original implementation. Figure \ref{fig:vnet} shows the network architecture with an input patch size of $64x64x64$.

The network has been trained using a patch size of $96^3$ and the random tumor distribution strategy (see \ref{sampling}). The maximum batch size due to memory constraints is 2.

\subsubsection{\textbf{3D-UNet}}

We use the original implementation with some minor modifications. Batch Normalization is changed for Group Normalization and, as in V-Net, we use 32 feature maps at the highest resolution. 

\begin{figure}[ht]
    \centering
    \includegraphics[width=1\textwidth]{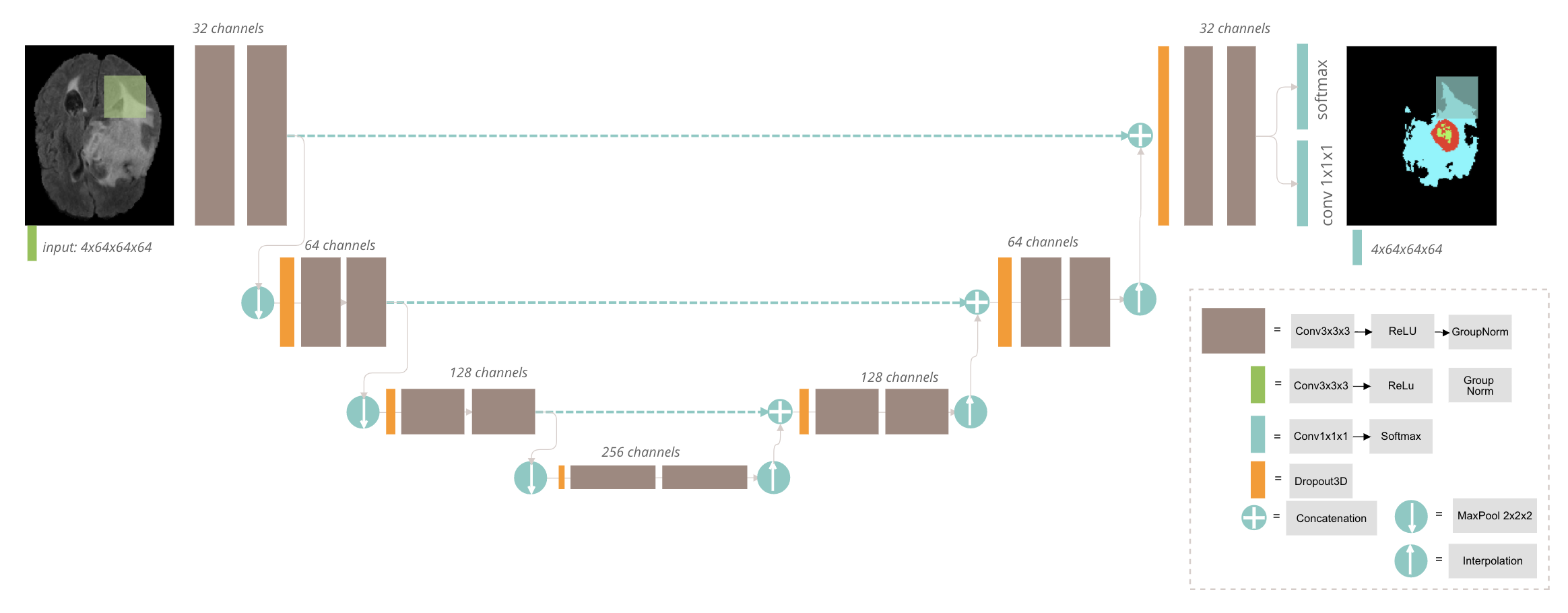}
    \caption{3D-Unet\cite{3dunet} architecture with Group Normalization, MaxPooling and Interpolation Upsampling and ReLU non-linearity} 
    \label{fig:3dunet}
\end{figure}

The network architecture is divided into symmetric Encoder and Decoder parts. The Encoder is composed of two convolutional blocks - with 3DConv + ReLu + GroupNorm structure. The downsampling is performed with $2^{3}$ Max-Pooling and the corresponding upsampling is performed with interpolation. All convolutional layers have kernel size $3^{3}$, except for the last one that has $1x1x1$ kernel and 4 feature maps as output. In this case, we use ReLu non-linearity and the skip-connections are joined with a concatenation step. The network outputs a four-channel segmentation map with the training labels as well as a softmax. The detailed architecture can be seen in Figure \ref{fig:3dunet}.

The \textit{Basic 3D-UNet} is trained with a patch size of $112^3$ and a batch size of 2.

\subsubsection{\textbf{Residual 3D-UNet}}

Expands the previous network with residual connections to allow having a deeper network with less risk of suffering from vanishing gradient. Adding to the residual blocks, the network also introduces some modifications w.r.t the basic 3D-UNet: (1) it uses element-wise sum to join the skip-connections, (2) it changes upsampling with interpolation for transposed convolutions and (3) it adds more depth to the network thanks to the resnet connections.

\begin{figure}[ht]
    \centering
    \includegraphics[width=1\textwidth]{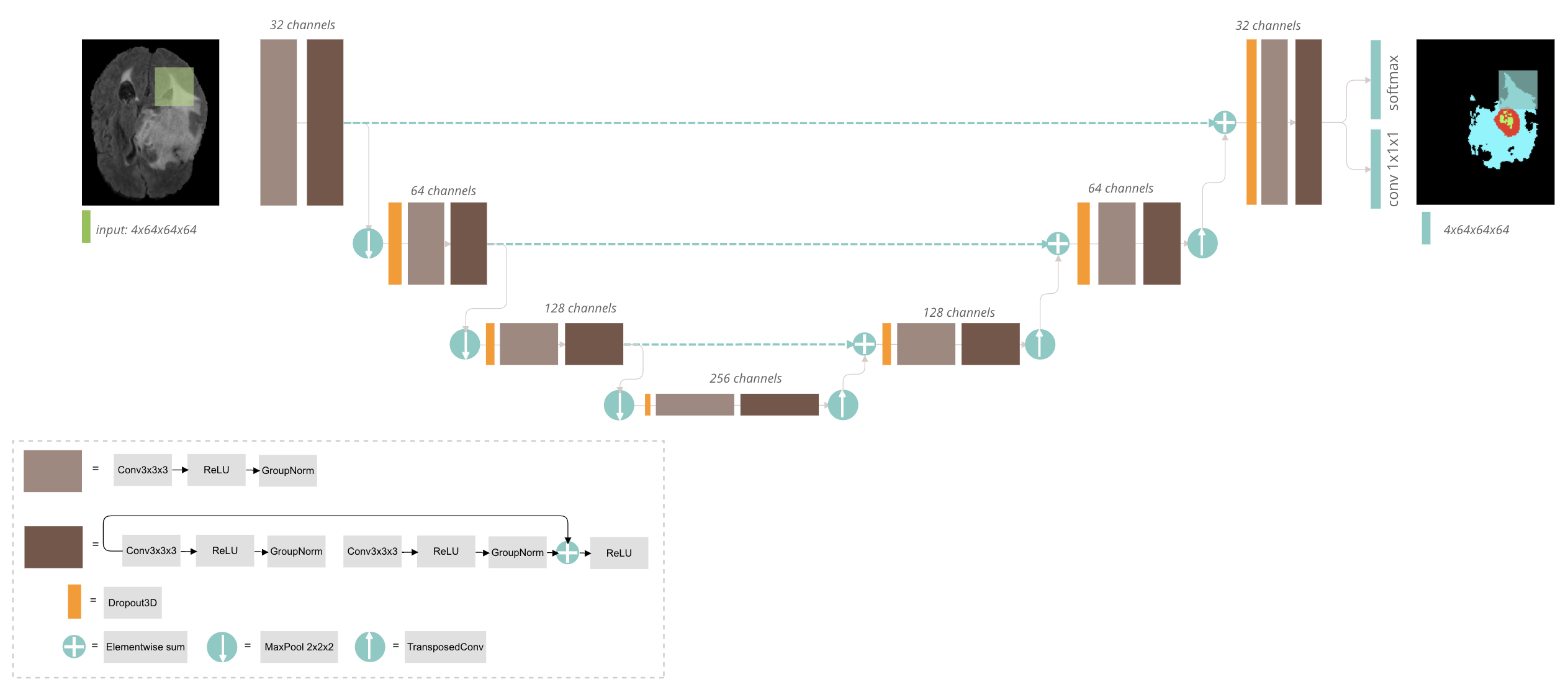}
    \caption{3D-Unet\cite{3dunet} architecture with RestNet blocks at each level, MaxPooling, TransposedConvolutions and ReLU non-linearity} 
    \label{fig:3dunet-resiudal}
\end{figure}

This network is trained following two different strategies. The first one, \textit{3D-UNet-residual} uses a patch size of $112^3$ and a batch size of 2 for the whole training, whereas \textit{3D-UNet-residual-multiscale} varies the sampling strategy so the network sees local and global information. For that, the first half of the training uses a patch size of $128^3$ with a batch size of 1. Then, the patch size is reduced to $112^3$ and the batch increased to 2.

\subsection{Post-Processing} \label{postprocessing}

In order to correct the appearance of false positives in the form of small and separated connected components, this work uses a post-processing step that keeps the two biggest connected components if their proportion is bigger than some threshold -obtained by analysing the training set. With this process, small connected components that may be false positives are removed but big enough components are kept as some of the subjects may have several tumors. 

Moreover, one of the biggest difficulties of this challenge is to provide an accurate segmentation of the smallest sub-region, ET, which is particularly difficult to segment in LGG patients, as almost 40\% have no enhancing tumor in the training set. In the evaluation step, BraTS awards a Dice score of 1 if a label is absent in both the ground truth and the prediction. Conversely, only a single false positive voxel in a patient where no enhancing tumor is present in the ground truth will result in a Dice score of 0. Therefore, some previous works \cite{no_newnet, two-stage} propose to replace enhancing tumor voxels for necrosis if the total number of enhancing voxels is smaller than some threshold, which is found for each experiment independently. However, we were not able to find a threshold that improved the performance as it helped for some subjects but made some other results worse.

\subsection{Uncertainty}
This year's BraTS includes a third task to evaluate the model uncertainty and reward methods with predictions that are: (a) confident when correct and (b) uncertain when incorrect. In this work, we model the voxel-wise uncertainty of our method at test time, using test time dropout (TTD) and test-time data augmentation (TTA) for epistemic and aleatoric uncertainty respectively.

We compute epistemic uncertainty as proposed in Gal et.al \cite{gal2015dropout}, who uses dropout as a Bayesian Approximation in order to simplify the task. Therefore, the idea is to use dropout both at training and testing time. The paper suggests to repeat the prediction a few hundred times with random dropout. Then, the final prediction is the average of all estimations and the uncertainty is modelled by computing the variance of the predictions. In this work, we perform $B=20$ iterations and use dropout with a 50\% probability to zero out a channel. The uncertainty map is estimated with the variance for each sub-region independently. Let $Y^{i} = \left \{ y_{1}^{i}, y_{2}^{i}...y_{B}^{i} \right \}$ be the vector that represents the i-th voxel's predicted labels, the voxel-wise uncertainty map, for each evaluation region, is obtained as the variance:

\begin{equation}
    var = \frac{1}{B} \sum_{b=1}^{B} (y_{b}^{i} - y_{mean}^{i})^{2}
\end{equation}

Uncertainty can also be estimated with the entropy, as \cite{wangUnc} showed. However, the entropy will provide a global measure instead of map for each sub-region. In this case, the voxel-wise uncertainty is calculated as:

\begin{equation}
    H(Y^{i}|X) \approx - \sum_{m=1}^{M} \hat{p}_{m}^{i}\ln(\hat{p}_{m}^{i})
\end{equation}

where $\hat{p}_{m}^{i}$ is the frequency of the m-th unique value in $Y^{i}$ and $X$ represents the input image.

To model aleatoric uncertainty we apply the same augmentation techniques from the training step plus random Gaussian noise, in order to add modifications not previously seen by the network. The final prediction and uncertainty maps are computed following the same strategies as in the epistemic uncertainty.

All that begin said, we hope to evaluate the model’s behaviour w.r.t to input and model variability by defining the several experiments:

\begin{itemize}
    \item Aleatoric Uncertainty: model aleatoric uncertainty with (1) TTA-variance, providing three uncertainty maps (ET, TC, WT) and (2) TTA-entropy, with one global map.
    
    \item Epistemic Uncertainty: model epistemic uncertainty with (1) TTD-variance, providing three uncertainty maps (ET, TC, WT) and (2) TTD-entropy, with one global map.
    
    \item Hybrid (Aleatoric + Epistemic) Uncertainty: model both aleatoric and epistemic uncertainty together with (1) TTD+TTA-variance, providing three uncertainty maps (ET, TC, WT) and (2) TTD+TTA-entropy, with one global map.
    
\end{itemize}

\section{Results}

The code\footnote{Github repository: https://github.com/imatge-upc/mri-braintumor-segmentation} has been implemented in Pytorch \cite{torch} and trained on the GPI\footnote{The Image and Video Processing Group (GPI) is a research group of the Signal Theory and Communications Department, Universitat Politècnica de Catalunya.} servers, based on 2 Intel(R) Xeon(R) @ 2.40GHz CPUs using 16GB RAM and a 12GB NVIDIA GPU, using BraTS 2020 training dataset. We report results on training, validation and test datasets. All results, prediction and uncertainty maps, are uploaded to the CBICA’s Image Processing Portal (IPP) for evaluation of Dice score, Hausdorff distance (95th percentile), sensitivity and specificity per each class. Specific uncertainty evaluation metrics are the ratio of filtered TN (FTN) and the ratio of filtered TP (FTP).

\subsection{Segmentation}

The principal metrics to evaluate the segmentation performance are the Dice Score, which is an overlap measure for pairwise comparison of segmentation mask X and ground truth Y:

\begin{equation}
    DSC = 2* \frac{\left | X \cap  Y \right |}{|X| + |Y|}   
\end{equation}

and the Hausdorff distance, which is the maximum distance of a set to the nearest point in the other set, defined as:

\begin{equation}
    D_{H}(X,Y) = max\left \{ sup_{x\epsilon X} \inf _{y\epsilon Y} d(x,y)) , sup_{y\epsilon Y} \inf _{x\epsilon X} d(x,y)) \right \}
\end{equation}

where sup represents the supremum and inf the infimum. In order to have more robust results and to avoid issues with noisy segmentation, the evaluation scheme uses the 95th percentile.

Tables \ref{tab:train} and \ref{tab:validation} show Dice and Hausdorff Distance (95th percentil) scores for training and validation sets respectively.

\begin{table}[ht]
    \caption{Segmentation Results on Training Dataset (369 cases).}\label{tab:train}
    \centering
    \begin{tabular}{l c c c c c c}
        \toprule
        \multirow{2}{*}{\textbf{Method}} & \multicolumn{3}{c}{\textbf{Dice}} & \multicolumn{3}{c}{\textbf{Hausdorff (mm)}} \\[1ex]
         & WT & TC & ET & WT & TC & ET \\
        \midrule
V-Net                       &	\textbf{0.87}	& 0.83	        & 0.74           & 10.19           & 12.89           &	35.96 \\ [1ex]
Basic 3D-UNet                      &	0.85	        & 0.84          & 0.76	         & \textbf{6.97}   & 10.13           & 28.23  \\ [1ex]
Residual 3D-UNet                   &	0.82	        & 0.82	        & 0.76	         & 8.56            & 12.11           & 28.93 \\ [1ex]
Residual 3D-UNet-multiscale        &	0.84	        & 0.84	        & 0.76	         & 7.43	           & 12.37           & \textbf{27.09} \\ [1ex]
Ensemble - mean                    &    0.85 	        & \textbf{0.85} & \textbf{0.77}  & 10.46	       & \textbf{6.90}   & 	29.03	 \\ [1ex]
        \bottomrule
    \end{tabular}
\end{table}

\begin{table}[ht]
    \caption{Segmentation Results on Validation Dataset (125 cases)}\label{tab:validation}
    \centering
    \begin{tabular}{l c c c c c c }
        \toprule
        \multirow{2}{*}{\textbf{Method}} & \multicolumn{3}{c}{\textbf{Dice}} & \multicolumn{3}{c}{\textbf{Hausdorff (mm)}} \\[1ex]
         & WT & TC & ET & WT & TC & ET\\
        \toprule
V-Net + post	                & \textbf{0.86}	 & 0.78	         & 0.69	          & 14.50	         & 16.15	         & 43.52	       \\ [1ex] 
Basic 3D-UNet +post	                & 0.81	         & 0.78	         & 0.67	          & 13.10	         & 14.01	         & 43.89	       \\ [1ex] 
Residual 3D-UNet + post         	& 0.81	         & 0.78          & 0.71	          & 11.85	         & 18.82	         & \textbf{34.97}\\ [1ex]	
Residual 3D-UNet-multiscale + post	& 0.83	         & 0.77	         & 0.72	          & 12.34	         & 13.11	         & 37.42	       \\ [1ex] 
Ensemble mean + post	            & 0.84	         & \textbf{0.79} & \textbf{0.72}  & 10.93	         & \textbf{12.24}    & 37.97	     \\ [1ex] 

        \bottomrule
    \end{tabular}
\end{table}

The model used with the test set is the Residual 3D-UNet-multiscale with post-processing. Table \ref{tab:test} shows the results in the training, validation and test sets for comparison.

\begin{table}[ht]
    \caption{Segmentation Results for model Residual 3D-UNet-multiscale + post on the three datasets}\label{tab:test}
    \centering
    \begin{tabular}{l c c c | c c c }
        \toprule
        \multirow{2}{*}{\textbf{Dataset}} & \multicolumn{3}{c|}{\textbf{Dice}} & \multicolumn{3}{c}{\textbf{Hausdorff (mm)}} \\[1ex]
         & WT & TC & ET & WT & TC & ET\\
        \toprule
            Train	& 0.84	  & 0.84	 & 0.76 & 7.43  & 12.37 & 27.09  \\ [1ex]
            Valid	& 0.83	  & 0.77	 & 0.72 & 12.34 & 13.11 & 37.42	 \\ [1ex] 
            Test	& 0.81	  & 0.82	 & 0.77 & 12.59 & 19.73 & 21.96	 \\ [1ex] 
        \bottomrule
    \end{tabular}
\end{table}

\begin{figure}[ht] 
    \centering
    \includegraphics[width=1\textwidth]{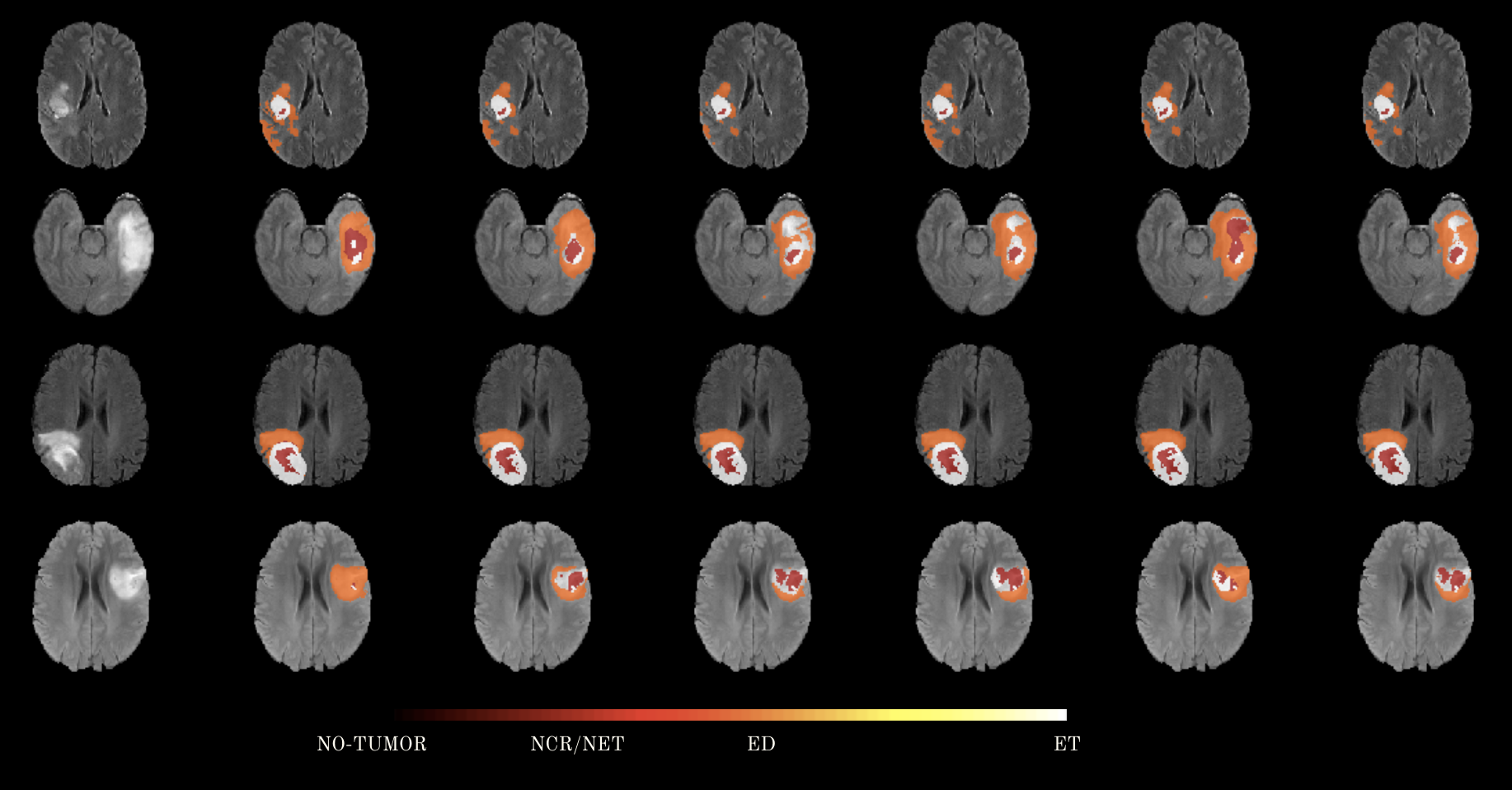}
    \caption{Training results on patients: 280, 010, 331 and 178 (top-bottom). Image order: (1) Flair (2) GT (3) Residual 3D-UNet-multiscale (4) Residual 3D-UNet (5) Basic 3D-UNet (6) V-Net (7) Ensemble mean }
    \label{fig:segmentation-results}
\end{figure}

All the proposed models are greatly penalized when no ET is present on the ground truth. In addition, the V-Net suffers more from false positives and 3D-UNet based models from false negatives. The excess of false positives may be caused due to the usage of small patches instead of using the whole volume, which provokes a variation in the proportion of healthy tissue against tumor regions. On the other hand, 3D-UNet models use bigger patch sizes and pooling layers instead of strided convolutions which may be the cause of having a larger number of false negatives. Increasing the patch size helps reduce false positives but it misses local information, which is reflected in label miss-classification on the region's boundaries. Figure \ref{fig:segmentation-results} shows a visual comparison of the models with a representation on the explained behaviours.

\subsection{Uncertainty}

BraTS requires to upload three uncertainty maps, one for each subregion (WT, TC, ET) together with the prediction map. Values must be normalized between 0-100 such that "0" represents the most certain prediction and "100" represents the most uncertain. The metrics used are the FTP ratio defined as $FTP = (TP_{100} - TP_{T}) / TP_{100}$, where T represents the threshold used to filter the more uncertain values. The ratio of filtered true negatives (FTN) is calculated in a similar manner. The integrated score will be calculated as follows:

\begin{equation}
    score = AUC_1 + (1-AUC_2) + (1-AUC_3). 
\end{equation}

From this point forward all experiments are performed on the model \textit{Residual 3D-UNet-multiscale}, as it is the one with more balanced results across the different regions. Table \ref{tab:validation_unc} shows the results for the epistemic, aleatoric and hybrid uncertainties when computed with entropy or variance. As a general overview, we can see that the AUC-Dice, which is computed by averaging the segmentation results for several thresholds that filter uncertain predictions, improves 2 to 3 points w.r.t the results obtained in the segmentation task ($WT: 0.8172$, $TC: 0.7664$, $ET: 0.7071 $). Although the metrics are not the same, it indicates that the model is more certain on the TP and less certain on FP and FN. Moreover, the AUC-Dice is higher when using entropy as the uncertainty measure.

Our results show that the model is more uncertain in LGG patients, particularly on epistemic uncertainty; meaning the model requires more data to achieve a more confident prediction. If we compare the behaviour between the uncertainty types, we see that (1) aleatoric focuses on the region boundaries, with small variations (2) epistemic improves results on the ET region but filters more TP and TN and(2) the hybrid approach achieves the best Dice-AUC results when using entropy as the uncertainty measurement.

\begin{table}[ht]
    \caption{Validation results on the \textit{Residual 3D-UNet-multiscale} for the followed approaches to estimate uncertainty.}\label{tab:validation_unc}
    \centering
    \begin{tabular}{l | c | c c c | c c c | c c c}
        \toprule
         \multirow{2}{*}{\textbf{Measure}} & \multirow{2}{*}{\textbf{Method}} & \multicolumn{3}{c|}{\textbf{Dice Score}} & \multicolumn{3}{c|}{\textbf{Ratio FTP}} & \multicolumn{3}{c}{\textbf{Ratio FTN}} \\ [1ex]
         & & WT & TC & ET & WT & TC & ET & WT & TC & ET \\
        \midrule
        \multirow{3}{*}{\textbf{Variance}}
        & TTA      & 0.83     	    & \textbf{0,77}	& 0,71	         & \textbf{0,05}  & \textbf{0,05}  & \textbf{0,04} & \textbf{9.0e-4}& \textbf{2.0e-4}  & \textbf{1.0e-4} \\ [1ex]
        & TTD      & 0.83	        & 0,76	        & \textbf{0,73}	 & 0,17	          & 0,16	       & 0,09	       & 2.4e-3 	      & 1.5e-3	      & 4.0e-4 \\ [1ex]
         & Hybrid  & \textbf{0,83}	& 0,76	        & 0,73	         & 0,18           & 0,16	       & 0,10	       & 3.6e-3           & 2.0e-3	      & 5.0e-4 \\ [1ex]
         \midrule
         \multirow{3}{*}{\textbf{Entropy}}
          &  TTA      & 0,83	        & 0,78	        & 0,71	        & 0,06	& \textbf{0,05}	& \textbf{0,04}	& \textbf{1.1e-3}&\textbf{4.7e-3}&\textbf{6.3e-3}  \\ [1ex]
          &   TTD     & 0,82	        & 0,78         	& 0,74	        & 0,15	& 0,13	        & 0,07	        & 2.1e-3	     & 8.2e-3	     & 1.22e-2 \\ [1ex]
          &  Hybrid   &\textbf{0,83}	& \textbf{0,79}	& \textbf{0,77}	& 0,15	& 0,12	        & 0,07	        & 3.0e-3	     & 1.01e-3	     & 1.39e-2 \\ [1ex]
            
        \bottomrule
        
    \end{tabular}
\end{table}

We participate in the challenge using test time augmentation (TTA) when uncertain values are computed using variance as it achieves the highest integrated scores per sub-region on the validation set. Table \ref{tab:validation_unc_final} shows the obtained results in both validation and test sets. The achieved integrated scores for validation are 0.93, 0.91 and 0.89 and for test 0.93, 0.93, 0.91 for WT, TC and ET respectively. We see a two point improvement on the ET and TC sub-regions for the test set.

\begin{table}[ht]
    \caption{Uncertainty Results for the \textit{Residual 3D-UNet-multiscale} model computed using TTA and variance for each sub-region independently. We show results for validation and test set for comparison }\label{tab:validation_unc_final}
    \centering
    \begin{tabular}{l c c c  c c c  c c c}
        \toprule
        \multirow{2}{*}{\textbf{Dataset}} & \multicolumn{3}{c}{\textbf{DICE AUC}} & \multicolumn{3}{c}{\textbf{FTP RATIO AUC}} & \multicolumn{3}{c}{\textbf{FTN RATIO AUC}} \\[1ex]
         & WT & TC & ET & WT & TC & ET & WT & TC & ET \\
        \toprule
        Valid & \textbf{0.8316}	&0.7715         &	0.7088       &	0.0449         &\textbf{0.0538}&\textbf{0.0380}&	\textbf{0.0009}     &	\textbf{0.0002}      &	\textbf{0.0001} \\ [1ex]
        Test  & 0.8299	        &\textbf{0.8124}&\textbf{0.7654} &	\textbf{0.0332}&0.0537         &0.0395         &	0.0020      &	0.0005      &	0.0003 \\ [1ex]
        \bottomrule
        
    \end{tabular}
\end{table}

\section{Discussion and Conclusions}

This work proposes a set of models based on two 3D-CNNs specialized in medical imaging, V-Net and 3D-UNet. As each of the trained models performs better in a particular tumor region, we define an ensemble of those models in order to increase the performance. Moreover, we analyze the implication of uncertainty estimation on the predicted segmentation in order to understand the reliability of the provided segmentation and identify challenging cases, but also as a means of improving the model accuracy by filtering uncertain voxels that should refer to wrong predictions. We use the Residual 3D-UNet-multiscale as our model to participate at the BraTS'20 challenge.

The best results in the validation set are obtained when creating an ensemble of the proposed models, as we can leverage the biases of each model, but are still far from the current state the art. These results may be caused by a bad training strategy where the sampling technique does not reflect the correct label distribution, thus providing more false detections. This is reflected more in the ET region as all models predict more tumor voxels of this label, which is greatly penalized when the ground truth does not contain it. In order to improve results, future work should try to provide a better representation of the labels, not just increase the patch size, but maybe let the network see both local and more global information.

Another potential problem is the model's simplicity. Although previous works achieve good results using a 3D-UNet, i.e. \cite{no_newnet}, adding more complexity to the network may help boost the performance. Therefore a possible line of work would be to extend the proposed models into a cascaded network, where each nested evaluation region --WT, TC and ET-- is learnt as a binary problem. Also, LGG subjects usually achieve lower accuracy on the prediction. In order to improve the results, we could research other post processing techniques and design them specifically to target each one of the glioma grades, as they may be differentiated by the sub-region distribution.

For uncertainty estimation, the work evaluates the usage of aleatoric, epistemic and a hybrid approach using the entropy as a global measure and variance to evaluate uncertainty on each evaluation region. In the provided results, it has been seen that using uncertainty information actually helps improve the accuracy of the network, achieving the best Dice Score (AUC, estimated from filtering uncertain voxels) when using the hybrid approach and entropy as the uncertainty measure. Our method achieves a score of 0.93, 0.93, 0.91 for WT, TC and ET respectively on the test set.

%
%

\begin{thebibliography}{8}
\bibliographystyle{splncs04}


\bibitem{brats}
B. H. Menze, A. Jakab, S. Bauer, J. Kalpathy-Cramer, K. Farahani, J. Kirby, et al.: "The Multimodal Brain Tumor Image Segmentation Benchmark (BRATS)", IEEE Transactions on Medical Imaging 34(10), 1993-2024 (2015) \doi{ 10.1109/TMI.2014.2377694}

\bibitem{advancing_brats}
S. Bakas, H. Akbari, A. Sotiras, M. Bilello, M. Rozycki, J.S. Kirby, et al.: "Advancing The Cancer Genome Atlas glioma MRI collections with expert segmentation labels and radiomic features", Nature Scientific Data, 4:170117 (2017) \doi{10.1038/sdata.2017.117}

\bibitem{bakas_identifying}
S. Bakas, M. Reyes, A. Jakab, S. Bauer, M. Rempfler, A. Crimi, et al.: "Identifying the Best Machine Learning Algorithms for Brain Tumor Segmentation, Progression Assessment, and Overall Survival Prediction in the BRATS Challenge", arXiv preprint arXiv:1811.02629 (2018)

\bibitem{segmentation_label_gbm}
S. Bakas, H. Akbari, A. Sotiras, M. Bilello, M. Rozycki, J. Kirby, et al., "Segmentation Labels and Radiomic Features for the Pre-operative Scans of the TCGA-GBM collection", The Cancer Imaging Archive, 2017. DOI: 10.7937/K9/TCIA.2017.KLXWJJ1Q

\bibitem{segmentation_label_lgg}
S. Bakas, H. Akbari, A. Sotiras, M. Bilello, M. Rozycki, J. Kirby, et al., "Segmentation Labels and Radiomic Features for the Pre-operative Scans of the TCGA-LGG collection", The Cancer Imaging Archive, 2017. DOI: 10.7937/K9/TCIA.2017.GJQ7R0EF

\bibitem{vnet}
Milletari, Fausto, Nassir Navab, and Seyed-Ahmad Ahmadi. ”V-net: Fully convo- lutional neural networks for volumetric medical image segmentation.” 3D Vision (3DV), 2016 Fourth International Conference on. IEEE, 2016.

\bibitem{epidemiology}
Morgan, L Lloyd: The epidemiology of glioma in adults: A "state of the science" review. Neuro-oncology vol.17 01-2015 \doi{10.1093/neuonc/nou358}

\bibitem{unc_categorization}
Armen Der Kiureghian and Ove Ditlevsen: Aleatory or epistemic? does it matter? Structural safety, 31(2):105–112, 2009.

\bibitem{montecarlo_dropout}
Yarin Gal and Zoubin Ghahramani: Dropout as a bayesian approximation: Representing model uncertainty in deep learning. arXiv preprint arXiv:1506.02142, 2015


\bibitem{deepmedic}
Konstantinos Kamnitsas, Christian Ledig, Virginia F.J. Newcombe, Joanna P. Simpson, Andrew D. Kane, David K. Menon, Daniel Rueckert, Ben Glocker:
Efficient multi-scale 3D CNN with fully connected CRF for accurate brain lesion segmentation, Medical Image Analysis, Volume 36, 2017, pages 61-78,
\doi{10.1016/j.media.2016.10.004}

\bibitem{Casamitjana1} Casamitjana, A., Puch, S., Aduriz, A., Vilaplana, V., "3D Convolutional Neural Networks for Brain Tumor Segmentation: a comparison of multi-resolution architectures". In: Brainlesion: Glioma, Multiple Sclerosis, Stroke and Traumatic Brain Injuries. BrainLes 2016. Lecture Notes in Computer Science, vol 10154. Springer, 2017.

\bibitem{ensembles}
Kamnitsas, K., Bai, W., Ferrante, E., McDonagh, S., Sinclair, M., Pawlowski, N: Ensembles of Multiple Models and Architectures for Robust Brain Tumour Segmentation in International MICCAI Brainlesion Workshop (Quebec, QC), 450–462 arXiv preprint arXiv:1711.01468, 2017

\bibitem{Casamitjana2} Casamitjana, A., Catà, M., Sánchez, I., Combalia, M., Vilaplana, V., "Cascaded V-Net Using ROI Masks for Brain Tumor Segmentation". In: Brainlesion: Glioma, Multiple Sclerosis, Stroke and Traumatic Brain Injuries. BrainLes 2017. Lecture Notes in Computer Science, vol 10670. Springer, 2018.

\bibitem{myronenko20183d}
Andriy Myronenko: 3D MRI brain tumor segmentation using autoencoder regularization. arXiv preprint arXiv:1810.11654, 2016

\bibitem{no_newnet}
Isensee, F., et al.: No new-net. International MICCAI Brainlesion Workshop, pp. 234–244. Springer (2018)

\bibitem{two-stage}
Jiang Z., Ding C., Liu M., Tao D. (2020) Two-Stage Cascaded U-Net: 1st Place Solution to BraTS Challenge 2019 Segmentation Task. In: Crimi A., Bakas S. (eds) Brainlesion: Glioma, Multiple Sclerosis, Stroke and Traumatic Brain Injuries. BrainLes 2019. Lecture Notes in Computer Science, vol 11992. Springer, Cham

\bibitem{tricks}
Zhao YX., Zhang YM., Liu CL. (2020) Bag of Tricks for 3D MRI Brain Tumor Segmentation. In: Crimi A., Bakas S. (eds) Brainlesion: Glioma, Multiple Sclerosis, Stroke and Traumatic Brain Injuries. BrainLes 2019. Lecture Notes in Computer Science, vol 11992. Springer, Cham

\bibitem{demistifying}
Natekar Parth, Kori Avinash, Krishnamurthi Ganapathy
AUTHOR=Natekar Parth, Kori Avinash, Krishnamurthi Ganapathy: Demystifying Brain Tumor Segmentation Networks: Interpretability and Uncertainty Analysis. Frontiers in Computational Neuroscience vol.14 page 6 \doi{10.3389/fncom.2020.00006}, 2020


\bibitem{wangUnc}
Wang G., Li W., Ourselin S. and Vercauteren T. Automatic Brain Tumor Segmentation Based on Cascaded Convolutional Neural Networks With Uncertainty Estimation. Frontiers in Computational Neuroscience vol.13 pages 56 \doi{10.3389/fncom.2019.00056}, 2019


\bibitem{McKinleyUnc}
McKinley R., Meier R., Wiest R. (2019) Ensembles of Densely-Connected CNNs with Label-Uncertainty for Brain Tumor Segmentation. In: Crimi A., Bakas S., Kuijf H., Keyvan F., Reyes M., van Walsum T. (eds) Brainlesion: Glioma, Multiple Sclerosis, Stroke and Traumatic Brain Injuries. BrainLes 2018. Lecture Notes in Computer Science, vol 11384. Springer

\bibitem{ulyanov2016instance}
Dmitry Ulyanov and Andrea Vedaldi and Victor Lempitsky. Instance Normalization: The Missing Ingredient for Fast Stylization. arXiv preprint arXiv:1607.08022, 2016

\bibitem{efficient_kamnitsas}
Kamnitsas, K., Ledig, C., Newcombe, V.F., Simpson, J.P., Kane, A.D., Menon, D.K., Rueckert, D., Glocker, B.: Efficient multi-scale 3D CNN with fully connected CRF for accurate brain lesion segmentation. Med. Image Anal. 36 (2017) 61–78

\bibitem{gal2015dropout}
Yarin Gal and Zoubin Ghahraman.Dropout as a Bayesian Approximation: Representing Model Uncertainty in Deep Learning. arXiv preprint arXiv:1506.02142, 2015

\bibitem{torch}
Paszke, Adam and Gross, Sam and Chintala, Soumith and Chanan, Gregory and Yang, Edward and DeVito, Zachary and Lin, Zeming and Desmaison, Alban and Antiga, Luca and Lerer, Adam. Automatic differentiation in PyTorch, NIPS-W 2017

\bibitem{unet}
Ronneberger, Olaf, Philipp Fischer, and Thomas Brox. ”U-net: Convolutional net- works for biomedical image segmentation.” MICCAI. Springer, 2015. \doi{10.1007/978-3-319-24574-4\_28}

\bibitem{gen_dice}
Sudre, C.H., Li, W., Vercauteren, T., Ourselin, S., Jorge Cardoso, M.Generalised Dice Overlap as a Deep Learning Loss Function for Highly Unbalanced Segmentations.  Lecture Notes in Computer Science 240-248, Springer International Publishing 2017. \doi{10.1007/978-3-319-67558-9\_28}

\bibitem{3dunet}
Özgün Çiçek and Ahmed Abdulkadir and Soeren S. Lienkamp and Thomas Brox and Olaf Ronneberger, "3D U-Net: Learning Dense Volumetric Segmentation from Sparse Annotation", arXiv preprint arXiv:1606.06650, 2016.

\bibitem{crum_dice_score}
W. R. {Crum} and O. {Camara} and D. L. G. {Hill}, "Generalized Overlap Measures for Evaluation and Validation in Medical Image Analysis", IEEE Transactions on Medical Imaging vol. 25, no. 11, pp. 1451–1461, 2006

\end{thebibliography}
%

\end{document}